# Gapped Electronic Structure of Epitaxial Stanene on InSb(111)

Cai-Zhi Xu[1,2,3], Yang-Hao Chan[4], Peng Chen[1,2,3], Xiaoxiong Wang[1,2,5], David Flötotto[1,2], Joseph Andrew Hlevyack[1,2], Guang Bian[6], Sung-Kwan Mo[3], Mei-Yin Chou[4,7,8] and Tai-Chang Chiang[1,2,8]

[1]Department of Physics, University of Illinois at Urbana-Champaign, Urbana, Illinois 61801, USA

[2]Frederick Seitz Materials Research Laboratory, University of Illinois at Urbana-Champaign, Urbana, Illinois 61801, USA

[3]Advanced Light Source, Lawrence Berkeley National Laboratory, Berkeley, California 94720, USA

[4]Institute of Atomic and Molecular Sciences, Academia Sinica, Taipei 10617, Taiwan

[5]College of Science, Nanjing University of Science and Technology, Nanjing 210094, China

[6]Department of Physics and Astronomy, University of Missouri, Columbia, Missouri 65211, USA

[7]School of Physics, Georgia Institute of Technology, Atlanta, Georgia 30332, USA

[8]Department of Physics, National Taiwan University, Taipei 10617, Taiwan

(Dated: Nov 21, 2017)

## Abstract

Stanene (single-layer grey tin), with an electronic structure akin to that of graphene but exhibiting a much larger spin-orbit gap, offers a promising platform for room-temperature electronics based on the quantum spin Hall (QSH) effect. This material has received much theoretical attention, but a suitable substrate for stanene growth that results in an overall gapped electronic structure has been elusive; a sizable gap is necessary for room-temperature applications. Here, we report a study of stanene epitaxially grown on the (111)B-face of indium antimonide (InSb). Angle-resolved photoemission spectroscopy (ARPES) measurements reveal a gap of 0.44 eV, in agreement with our first-principles calculations. The results indicate that stanene on InSb(111) is a strong contender for electronic QSH applications.

Since the discovery of graphene [1], two-dimensional (2D) materials have attracted much attention for their potentially novel properties relevant to electronic applications [2–6]. Of special interest are the analogs of graphene made of the other elements in the same column of the periodic table, which include silicene, germanene, and stanene [4,5]. All of them are predicted to be quantum spin Hall (QSH) systems with a band gap arising from spin-orbit coupling [7–11]. This gap makes these materials electrically insulating atomic sheets, but the edges of the sheets support 1D topological edge states that can carry robust currents with an inherent spin polarization well suited for spin signal processing [12,13]. For graphene, the strength of spin-orbit coupling is extremely weak; its gap is negligible (0.02 meV) [7] compared to the thermal energy $k_BT$ = 25 meV at room temperature, where the edge states would be electrically shorted out and useless as conduction channels. Moving down the periodic table, the spin-orbit gap increases with increasing atomic number [8,9], and freestanding stanene is predicated to have a gap of ~100 meV, which would allow for room-temperature applications of the topological edge states [8,10]. Theory further predicts that the gap in stanene can be enhanced to ~300 meV by adsorption of halogen atoms [10], which is an order of magnitude larger than the QSH gaps in all materials that have been experimentally established thus far [14–16]. Stanene is also of interest for its theoretically predicted enhanced thermoelectricity [17], topological superconductivity [18], and quantum anomalous Hall behavior [19]. Unfortunately, experimental fabrication of high-quality stanene with the desired electronic structure has proven to be difficult. Unlike graphene/graphite characterized by $sp^2$ bonding, grey Sn is characterized by $sp^3$ bonding. As a result, the lattice structure of stanene is not flat like graphene, but buckled as shown schematically in Fig. 1(a). This buckled structure contains dangling bonds and is naturally unstable against 3D cluster formation. The buckling, however, does not affect the surface Brillouin zone shown in Fig. 1(b).

Prior studies have shown that stanene films can be prepared by depositing Sn on selected substrates including $Bi_2Te_3$, Sb(111), and Bi(111) [20–22]. Presumably, bonding to the substrate counters the effects of the dangling bonds and stabilizes the formation of stanene. However, these systems are all metallic because of the presence of metallic (topological) surface states from the substrate or the semimetallic nature of the substrate material; as a result, electrically isolated QSH channels are not supported. Several semiconductors with sizable gaps, including CdTe [10], BN, InSb [23], PbTe, SrTe, BaSe and BaTe [24], are potential candidate substrate materials for growing stanene and stanene further modified by surface adsorption [10,24]. Surface adsorption could provide additional structural stabilization, but the chemical interaction could cause other undesirable effects. Each case will need careful evaluation by experimentation.

In the present work, we focus on InSb as a substrate material for growing stanene. InSb is an especially favorable case because its lattice is well matched to that of grey Sn. If the In and Sb atoms in InSb were all replaced by Sn atoms, the resulting structure would be just grey Sn with a lattice very slightly compressed by 0.14%. Experimentally, multilayer grey-Sn films have been successfully grown on InSb [25,26]. Here, using reflection high-energy electron diffraction (HREED) and angle-resolved photoemission spectroscopy (ARPES), we show that a high-quality single Sn layer can be grown on the (111)B-face of InSb (the B face is Sb-terminated, while the A face is In-terminated). Furthermore, the system exhibits a large gap of 0.44 eV; this gap size is consistent with our first-principles calculations. With increasing film thickness in the experiment, the system transforms from a gapped insulator to a semimetal already at just two layers. In the bulk limit, unstrained grey-Sn is a semimetal with a single-point Fermi surface at the zone center. The large gap of stanene relative to the bulk as seen in our experiment can be understood as a result of quantum confinement; it is a feature of importance relevant to QSH applications.

Our experiment was performed at the Advanced Light Source, Lawrence Berkeley National Laboratory, using the HERS endstation at Beamline 10.0.1. Substrates of InSb(111)B were cleaned by repeated cycles of Ar ion sputtering and subsequent annealing at 400 °C in a molecular-beam-epitaxy (MBE) chamber attached to the beamline; a sharp RHEED pattern characteristic of the 3×3 surface reconstruction of InSb was observed afterwards as shown in Fig. 1(d). A Knudsen cell was used to evaporate Sn at a rate of ~3 minutes per layer onto the substrate surface maintained at room temperature. The RHEED intensity exhibited oscillations during Sn deposition as shown in Fig. 1(c), which suggested a layer-by-layer growth mode. The oscillation period, corresponding to the deposition of a single layer, offered a means for precise layer thickness control. After the deposition of a single layer of Sn, the 3×3 RHEED pattern of InSb was fully converted into a 1×1 pattern of stanene as shown in Fig. 1(e). The spacings between the main (non-fractional-order) diffraction streaks remained unchanged in going from 3×3 to 1×1. Thus, the in-plane lattice constant of the deposited stanene layer is the same as that of the InSb substrate (4.58 Å). ARPES measurements were performed with the sample maintained at 40 K using *s*-polarized light and a Scienta R4000 analyzer; the energy and angular resolutions were 20 meV and 0.1°, respectively.

The photoemission spectrum in Fig. 1(f) was taken with 80-eV photons. The measured intensities of the 4*d* core levels of In, Sn, and Sb are consistent with a single layer of Sn on top of an InSb substrate. Doping of the surface with K, when needed, was performed with a SAES Getter dispenser.

First-principles calculations were performed using the projector augmented wave method [27,28] implemented in the Vienna ab initio package (VASP) [29,30]. A plane wave basis set with a cutoff energy of 400 eV and a Monkhorst-Pack *k*-space mesh of 12x12x1 were used. The InSb substrate was simulated by a 6-layer InSb slab. The atomic positions of the top three atomic layers including the stanene were determined by energy minimization within the local-density approximation (LDA) [31]. A vacuum gap of 15 Å between periodic slabs was used in the modeling. The optimized structure is indicated in Fig. 2(e). The down-buckled Sn atoms sit on top of the surface Sb sites of InSb with a vertical spacing of 2.85 Å. The electronic band structure was calculated using the LDA + U method. The Hubbard U parameters [32] for the Sn *p*-, In *p*-, and Sb *s*-orbitals were set to -3.5, 4, and -9, respectively, in order to yield the correct band ordering as reported in previous experimental [33,34] and theoretical studies [35].

Results of ARPES band mapping of the clean InSb substrate using 42-eV photons along the two high-symmetry directions $\overline{\Gamma K}$ and $\overline{\Gamma M}$ are shown in Fig. 2(a); to highlight the bands, the second derivative [36] of the ARPES map is shown in Fig. 2(c). The two top valence bands, which merge at the zone center, correspond to the heavy-hole (HH) and light-hole (LH) bands of InSb, respectively. The feature below corresponds to the split-off band. Figures 2(b) and 2(d) show the ARPES map and its second derivative, respectively, taken from stanene-covered InSb. The Fermi level position was determined from the Fermi edge of a piece of tantalum metal in direct contact with the sample. After the deposition of stanene, the valence band top shifts down to a binding energy of 0.44 eV with no other clearly visible ARPES features above, indicating a substantial energy gap in this system.

For comparison, the calculated band structure of stanene on InSb is shown in Fig. 2(f). The size of each dot for the band structure plot is chosen to correspond to the charge within the stanene layer plus the two top InSb layers. This should roughly correspond to the ARPES intensity because of the short ARPES probing depth, although the ARPES intensity also depends sensitively on the photoemission cross section of each state. The Fermi level is assumed to be located at the conduction band minimum, which would correspond to a lightly *n*-doped substrate (see below for further justification). Evidently, the theoretical valence band features are in good agreement with the experiment around the zone center, except that the calculated gap of 0.33 eV is somewhat smaller than the experimental value; the discrepancy is within the usual uncertainty range of this type of calculations. Theory also shows two relatively slow-dispersing bands near the zone boundaries that are not detected in the experiment. Similar disagreement was reported in a prior study, but the exact reason remains unknown [20]. It is possible that the cross section for these states (mostly derived from the $p_z$ orbital of Sn) is just unfavorable for ARPES detection with an *s*-polarization configuration.

In any case, the results confirm that stanene on InSb is an insulator with a substantial gap. For comparison, the band gap of freestanding stanene is only ~0.1 eV based on theory [10]. The enhanced band gap of stanene on InSb is likely a result of the interaction between the stanene and the substrate, just like the case of adsorbate-induced gap widening mentioned earlier. To further pin down the gap value, we have employed potassium (K) doping of the sample surface in order to bring out the conduction bands in ARPES measurements. Figures 3(a) and 3(b) show close-up ARPES maps of the electronic structure of stanene on InSb before and after K doping, respectively; the corresponding second-derivative maps are shown in Figs. 3(c) and 3(d). After K-deposition, the system becomes heavily *n*-doped, and a V-shaped conduction band around the zone center becomes evident in Fig. 3(d). The V-shape implies a very small effective mass and a very high electron group velocity close to the zone center. The gap determined from Fig. 3(d) is 0.44 eV. A close inspection of Fig. 3(c) around the zone center reveals a very faint single-point-like emission feature at the Fermi level. This can be identified as the conduction band bottom before K-doping; thus, the sample is slightly *n*-doped. The band gap before K-doping, measured from the data in Fig. 3(c), is 0.44 eV, same as that after K-doping. Figure 3(e) shows the calculated band diagram shifted downward in energy to indicate the occupation of the conduction bands by the K-doping. Further insight of the electronic structure of the stanene-InSb system is gleaned from the plot presented in Fig. 3(f) of the charge densities of the states at the conduction band bottom and the valence band top projected onto each atomic layer. The charge density associated with the valence band top is predominantly confined within the stanene layer ($p_{x,y}$-orbitals of stanene), while that for the conduction band bottom involves both the stanene and the InSb substrate (*s*-orbitals from both the stanene and InSb). The results indicate a strong electronic interaction or bonding across the interface through the conduction band states.

The evolution of the electronic structure from stanene (2D limit) to bulk grey Sn (3D limit) is of interest. Figure 4(a) shows ARPES maps taken along $\bar{\Gamma}\bar{K}$ and $\bar{\Gamma}\bar{M}$ and the corresponding second-derivative maps for grey-Sn films with thicknesses of 1, 2, 4, and 6 layers. The corresponding theoretical results assuming a 6-layer InSb substrate are shown in Fig. 4(b). The single Sn layer case with a gap of 0.44 eV has been discussed earlier. With the addition of a Sn layer, the gap vanishes according to theory, but the surface weights of the lower conduction bands are very small, and there is very little ARPES intensity in the data. The Sn film remains largely a semiconductor but with a much reduced gap. For 4 Sn layers, the valence band top moves further up and almost touches the Fermi level. Further increasing the Sn film thickness to 6 layers results in only small changes along the same trend. Evidently, the system is already very close to the bulk limit at Sn thicknesses of ~4-6 layers. The trend of gap evolution in going from the 3D to the 2D limit can be understood as a result of quantum confinement. In the bulk grey-Sn limit, the valence and conduction bands should overlap, but topological effects reduce the overlap to just one point, and the system is a 2D Dirac semimetal [37]. As the film thickness reduces, the widths of the Sn conduction and valence bands decrease, leading to band separation or gap opening. This is clear from the evolution of the Sn-dominated valence band top and the Sn-dominated conduction band denoted CB' in Fig. 4(b). Because of hybridization between the Sn and InSb conduction band states, the Sn CB' state is accompanied by a bundle of InSb conduction band states, which fill up the gap for the 2- and 4-layer cases in Fig. 4(b). At the single-layer limit, the CB' state has moved to a much higher energy, and the accompanying hybridized InSb states become separated from the valence band top, resulting in a gap of 0.44 eV.

In summary, we have successfully grown high-quality stanene films on the (111)B-face of InSb, a system of interest for possible room-temperature QSH applications. ARPES measurements from the pristine films and K-doped films demonstrate a large gap of 0.44 eV at the zone center. The major features of the bands near the zone center, including a very small effective mass and a very large group velocity near the conduction band minimum, are in good accord with first-principles calculations. The measured gap of 0.44 eV is substantially larger than that expected for freestanding stanene, which is a surprising but welcoming finding. The reason for the difference is electronic coupling between the Sn and InSb conduction band states, as demonstrated by a comparison between theoretical modeling and experiment for grey-Sn films of various thicknesses that bridge the 2D and 3D limits. Only the single-layer case exhibits a gap. At two layers, the gap is already filled in by InSb conduction band states; these substrate states would act as an electrical short, thus rendering the system unsuitable for edge electronics. The gap for stanene as reported herein is much larger than the thermal energy $k_BT = 25$ meV at room temperature. We conclude that the stanene film on InSb investigated in this study is a promising candidate for QSH applications at room temperature and higher.

*Acknowledgments* This work is supported by the U.S. Department of Energy (DOE), Office of Science (OS), Office of Basic Energy Sciences, Division of Materials Science and Engineering, under Grant No. DE-FG02-07ER46383 (T.C.C.). The Advanced Light Source is supported by the Office of Basic Energy Sciences of the U.S. DOE under Contract No. DE-AC02-05CH11231. C.Z.X. received partial support from the ALS Doctoral Fellowship in Residence during the experiment. M.Y.C. is supported by the U.S. National Science Foundation under Grant No. EFMA-1542747. Y.H.C. is supported by a Thematic Project at Academia Sinica. X.X.W. is supported by the National Science Foundation of China under Grant No. 11204133 and the Fundamental Research Funds for the Central Universities under Grant No. 30917011338. D.F. is supported by the Deutsche Forschungsgemeinschaft (FL 974/1-1).

**FIG. 1.** (a) Top and side views of the atomic structure of stanene. (b) Brillouin zone of stanene. (c) RHEED intensity as a function of time of Sn growth on InSb(111). The blue arrows mark when each layer of Sn is formed. (d) RHEED pattern of the InSb(111) substrate before deposition of Sn, which shows a 3×3 reconstruction. (e) RHEED pattern after one layer of Sn is deposited to form stanene. (f) Photoemission spectrum taken from stanene on InSb(111). The peaks correspond to the 4$d$ core levels of In, Sn, and Sb as labeled.

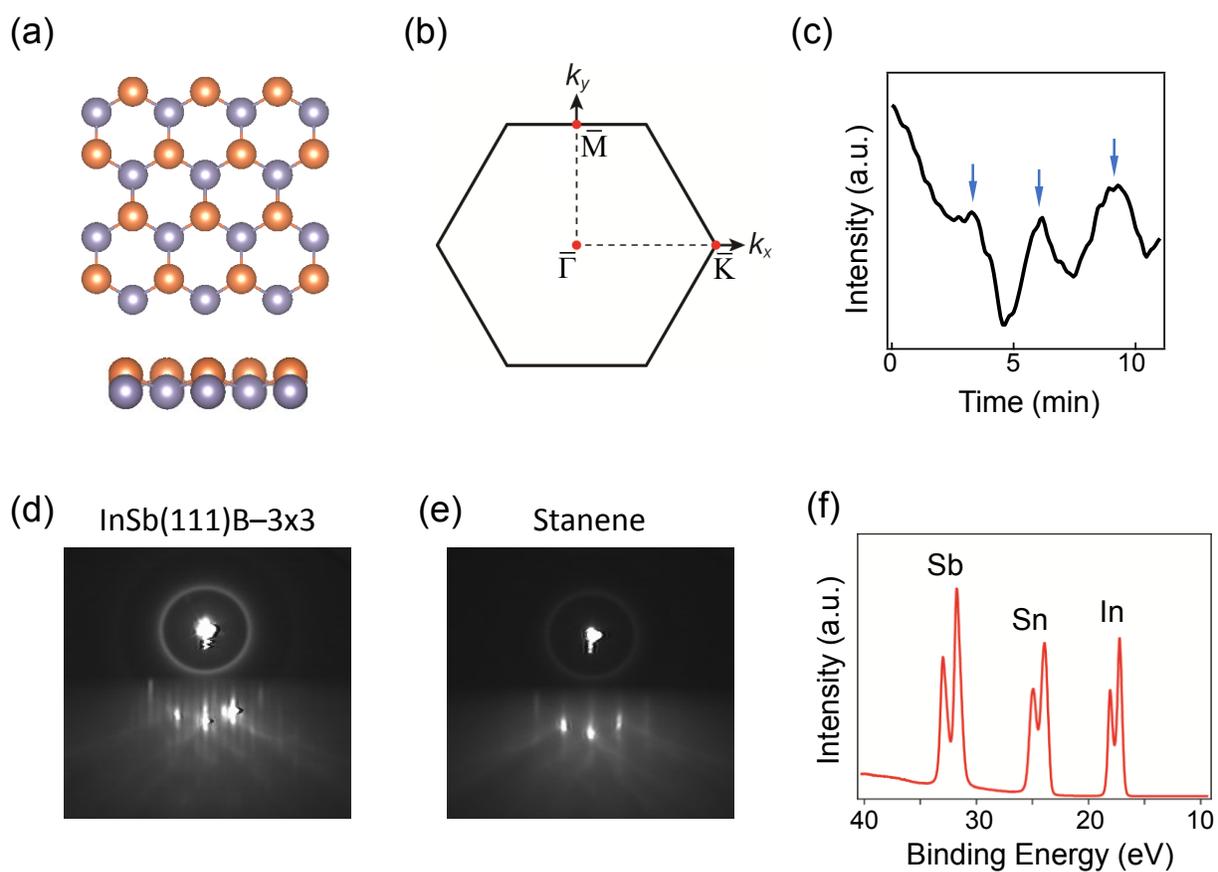

**FIG. 2.** (a) ARPES map for InSb(111) along $\overline{\Gamma K}$ and $\overline{\Gamma M}$. The heavy hole (HH) and light hole (LH) bands are indicated. (b) ARPES map for stanene-covered InSb(111). (c) Second-derive map derived from (a). (d) Second derivative map derived from (b). (e) Side view of the atomic structure of stanene-covered InSb(111). Only the top two layers of InSb are shown. (f) Calculated band structure for stanene on top of 6 layers of InSb(111). The conduction band (CB) and valence band (VB) are indicated. The dot size corresponds to the charge associated with each state within the stanene layer and the two top InSb layers. The Fermi level is set to the conduction band bottom.

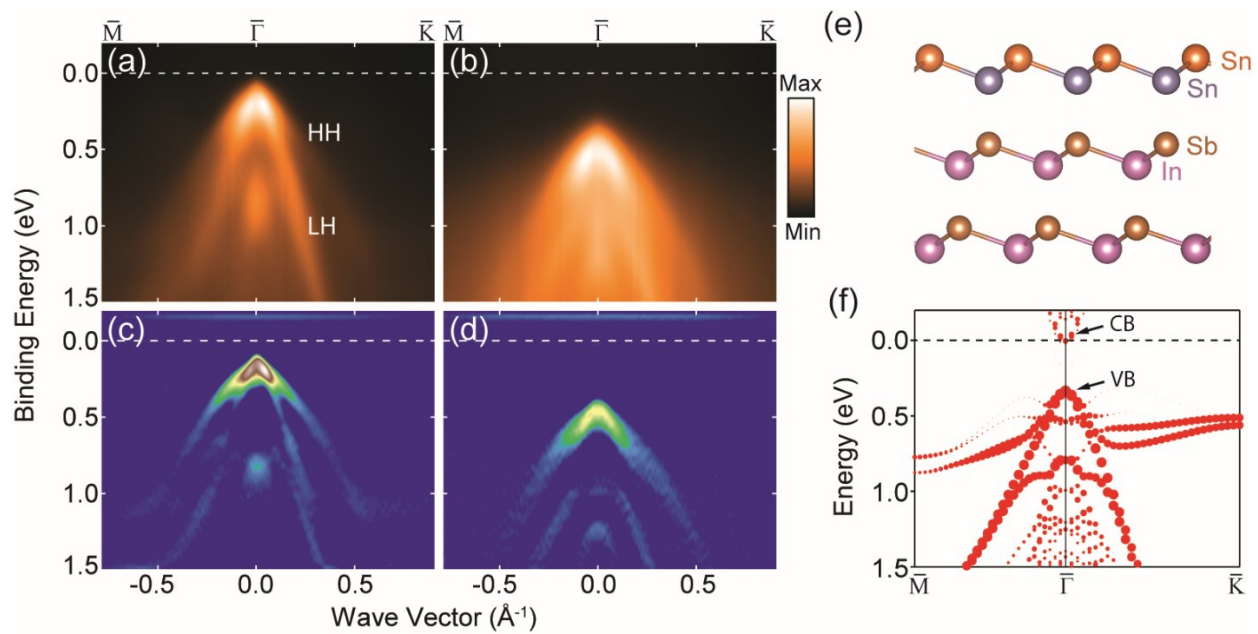

**FIG. 3.** (a) ARPES map of stanene on InSb. (b) ARPES map after K doping. The CB and VB are labeled. (c) Second-derivative map derived from (a). (d) Second-derivative map derived from (b). (e) Calculated band structure weighted by the surface charge. The Fermi level is adjusted for a close correspondence to the data after K doping. (f) Charge densities corresponding to the VB and CB at the zone center evaluated at each atomic layer for the stanene-covered InSb sample.

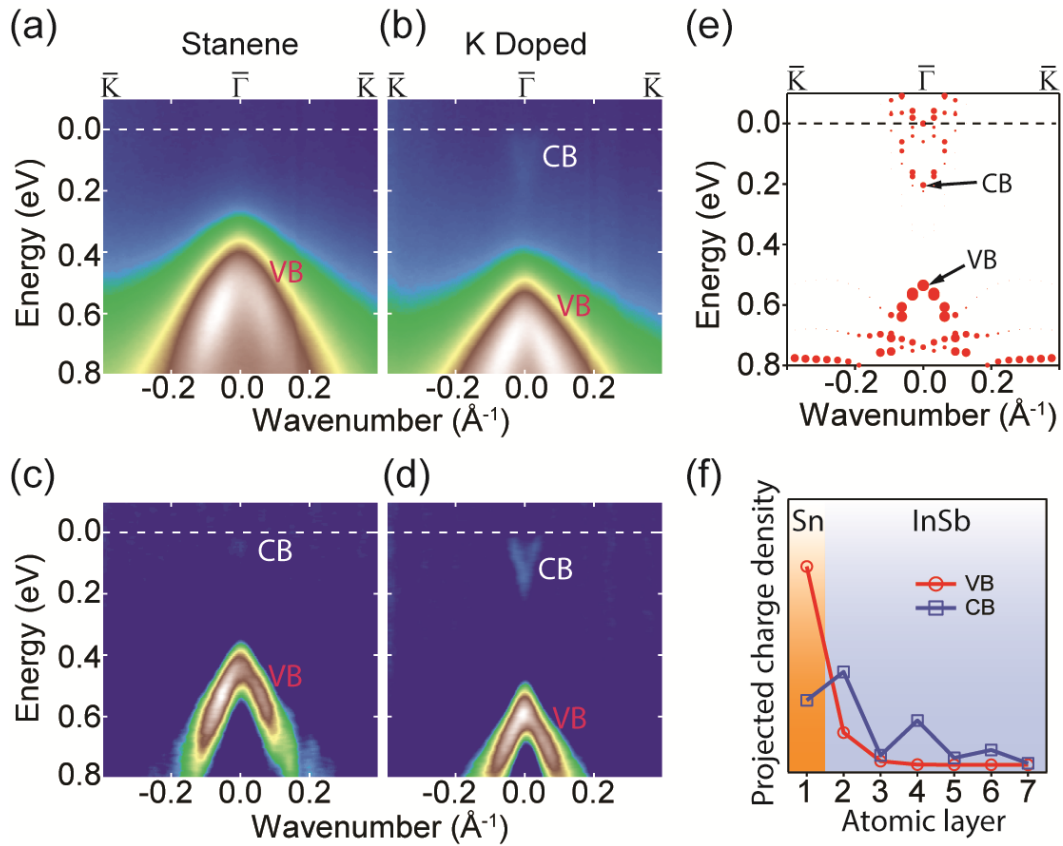

**FIG. 4.** (a) Top row: ARPES maps for 1-, 2-, 4-, and 6-layer grey Sn on InSb. Bottom row: corresponding second-derivative maps. (b) Corresponding calculated band structures. The size of each dot corresponds to the surface weight of the charge density for each state. CB' denotes a conduction band dominated by the Sn states. The other CB states with much less surface weight are mostly associated with InSb. The Fermi level in each case is adjusted to match the data.

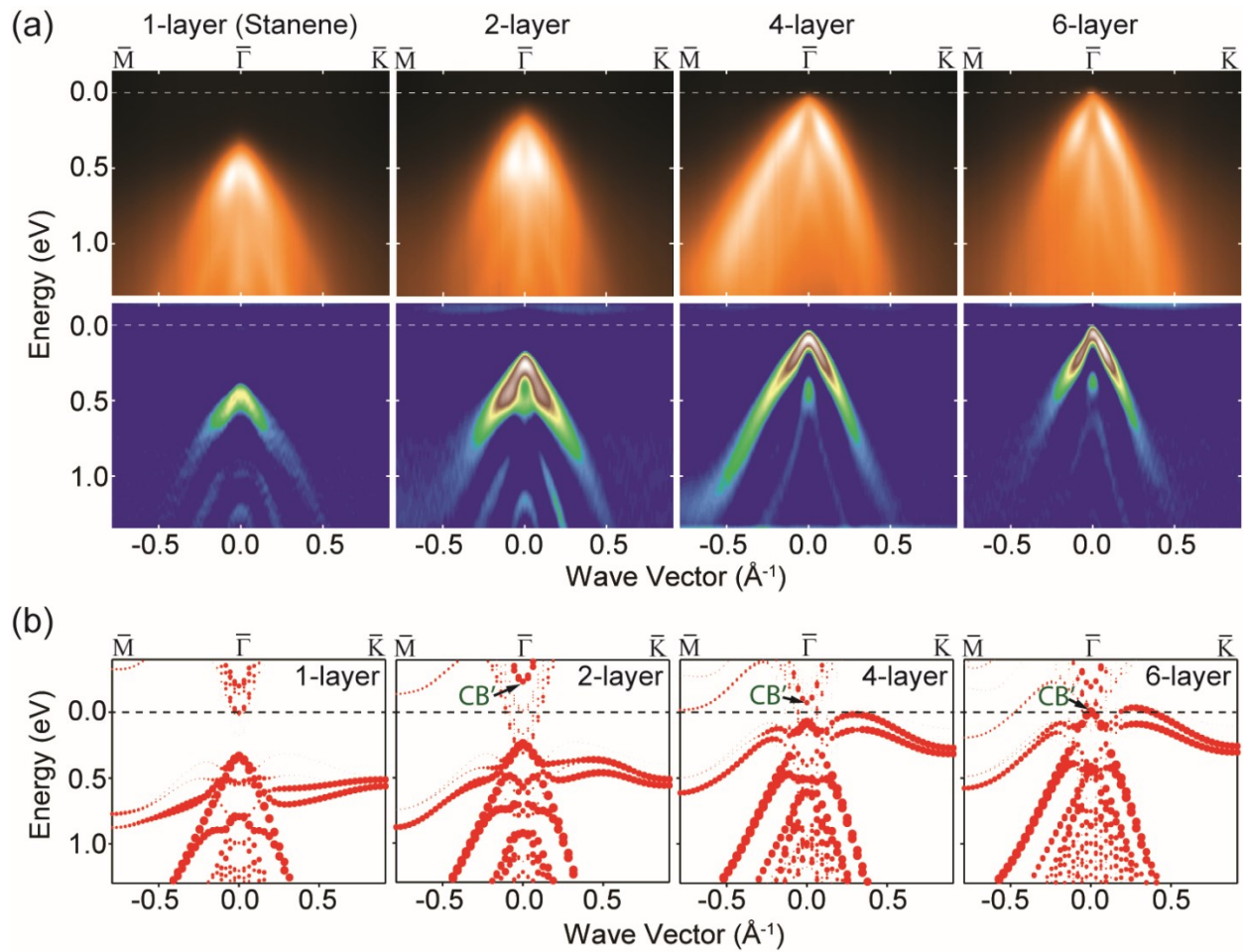